\documentclass[10pt]{article}
\usepackage{amsfonts}
\usepackage{amssymb}
\usepackage{amstext}

\usepackage{theorem}
\newtheorem{teo}{}
\newtheorem{theorem}[teo]{Theorem}
\newtheorem{proposition}[teo]{Proposition}
\newtheorem{corollary}[teo]{Corollary}
\newtheorem{definition}[teo]{Definition}
\newtheorem{lemma}[teo]{Lemma}

\newtheorem{remark}[teo]{Remark}

\newenvironment{proof}
	{\par {\it Proof:}}
 	{\hfill $\square$ \medskip}

\newcommand{\slim}{\mathop{\mbox{\rm s-lim}}}
\def\ds{\displaystyle}
\def\b{\begin{equation}}
\def\e{\end{equation}}
\def\mc{\mathcal}
\def\mb{\mathbb}
\def\N{{\mb N}}
\def\R{{\mb R}}
\def\C{{\mb C}}
\def\Z{{\mb Z}}
\def\H{{\mc H}}
\def\D{{\mc D}}
\def\G{{\mc G}}
\def\H{{\mc H}}
\def\I{{\mc I}}
\def\K{{\mc K}}
\def\L{{\mc L}}
\def\LL{\Lambda}
\def\RR{{\mc R}}
\def\l{\lambda}
\def\om{\omega}

\begin{document}

\title{Intrinsic randomness of unstable dynamics\\ and Sz.-Nagy-Foia\c s dilation theory}

\author{Fernando G\'omez}

\maketitle

\begin{center}
{\small
Dpto. de An\'alisis Matem\'atico, Universidad de Valladolid\\
Facultad de Ciencias, Prado de la Magdalena, s.n.\\
47005 Valladolid, Spain\\
e-mail: {\tt fgcubill@am.uva.es}
}
\end{center}

\begin{abstract}
Misra, Prigogine and Courbage (MPC) demonstrated the possibility of obtaining stochastic Markov processes from deterministic dynamics simply through a "change of representation" which involves no loss of information provided the dynamical system under consideration has a suitably high degree of instability of motion. From a mathematical point of view, MPC theory is a theory of positivity preserving quasi-affine  transformations that intertwine the unitary groups associated with deterministic dynamics to contraction semigroups associated with stochastic Markov processes.
In this work, dropping the positivity condition, a characterization of the contraction semigroups induced by quasi-affine transformations, the structure of the unitary groups admitting such intertwining relations and a prototype for the quasi-affinities are given on the basis of the Sz.-Nagy-Foia\c s dilation theory.
The results are applied to MPC theory in the context of statistical mechanics.
\end{abstract}




\section{Introduction}

The conventional topological approach to the study of classical dynamical systems is based on trajectories in the phase space $\Omega$ describing the point dynamics by a family $S_t$ of endomorphisms or automorphisms of $\Omega$, namely, the time evolution
$$
\om_0\mapsto \om_t:=S_t\om_0
$$
of single points $\om_0\in\Omega$, where $t\in\R$ or $t\in\R^+:=[0,\infty)$ for flows and $t\in\Z$ or $t\in\Z^+:=\{n\in\Z:n\geq0\}$ for cascades. 
For systems presenting strong instabilities of motion trajectories lose operational meaning and dynamics is usually formulated in terms of the motion of distribution functions.
In the probabilistic approach, extensively used in statistical mechanics and ergodic theory, trajectories are replaced by the study of the corresponding Koopman and/or Frobenius--Perron operators \cite{LM94}, which describe, respectively, the evolution of the observables and the probability densities of the system.
In the Hilbert space $L^2=L^2(\Omega,{\mathcal A},\mu)$ of square integrable functions on $\Omega$ --with respect to the reference $\sigma$-algebra ${\mathcal A}$ and measure $\mu$-- the Koopman operator $V_t$ and its $L^2$-adjoint, the Frobenius--Perron operator $U_t$, are defined as
$$
V_tf(\om):= f(S_t\om)\,,\qquad  (U_t\rho,f)=(\rho,V_tf)\,,
$$
where $(\rho,f)=\int_\Omega \rho(\om)\,f(\om)\,d\mu(\om)$ is the expectation value of the observable $f$ in the density $\rho$.
Reversible (automorphic) dynamics are then described in terms of a group of unitary operators $\{U_t\}$ acting on $L^2$.
In a similar way, for quantum dynamical systems the 
evolution of wave-functions or density operators is determined by the unitary group $U_t=e^{itH}$ generated by the Hamiltonian $H$.
In contrast, systems qualified by irreversible undirectness of the evolution are associated with contraction semigroups $\{W_t\}$, the time parameter $t$ taking positive (integer or real) values if evolution is directed towards the future.
Typical examples include heat equation, Boltzmann equation and stationary Markov processes.

The problem of reconciling the apparent irreversible behavior of (macroscopic) systems with the reversible nature of fundamental microscopic laws of physics, including both classical and quantum mechanics, is far from being completely solved. Experimental and numerical results for microscopic irreversibility have been recently published \cite{PG05}. These examples show that microscopic irreversibility is associated to chaotic behavior and does not require that dynamical equations violate time-reversal symmetry or that the system be coupled to a source of external noise.
In the late 1970's Misra, Prigogine and Courbage (MPC) \cite{M78,MPC79,GMC81,CP83} already discussed the question of the dynamical meaning of the second law of thermodynamics at microscopic level. MPC approach expresses irreversibility in terms of the existence of Lyapounov operators --i.e., observables varying monotonically in time-- and shows its close links with the inherent randomness of the system and its dynamical instability --for instance, mixing property is necessary--.
MPC {\it intrinsic randomness} is based on the existence of non-unitary (invertible) similarity transformations $\LL$  (called {\it quasi-affinities} in what follows) relating unitary dynamical groups $\{U_t\}$ with Markovian evolution semigroups $\{W_t\}$
through an intertwining relation of the form:
\b\label{Int.Rel}
W_t\LL=\LL U_t\,,\quad (t\geq0)\,.
\e
In contrast with ``coarse-grained" descriptions, relation (\ref{Int.Rel}) involves no loss of information and derived Markovian semigroups are not related to local point transformations in state space \cite{MP83,SAT94}. Trajectories lose then operational meaning and the above extended distributional framework of dynamics must be considered. 
On the other hand, contrary to ``open-system" evolution, where irreversible behavior is due to its interaction with environment, MPC theory refers to irreversible behavior originating in the own dynamics of the system.  

Following a suggestion by Misra \cite{M78}, intrinsically random unitary evolutions $\{U_t\}$ have been qualified by the existence of an {\it internal time operator} $T$, a self-adjoint operator satisfying \cite{G1}:
$$
U_{-t} T U_t=T+tI\,.
$$
The operator $T$ allows the attribution of an average age to each state $\rho$ which keeps step with the external clock time $t$ for the evolved state $U_t\rho$.
The transformation $\LL$ is then an operator function of the internal time $T$.
Further work has been done studying the connections between deterministic dynamics and probabilistic processes, but the question of intertwining by a quasi-affinity is not yet well enough understood --see \cite{B04} and references therein--.

This work deals with the intertwining relation (\ref{Int.Rel}) on the basis of the Sz.-Nagy-Foia\c s dilation theory \cite{NAGY-FOIAS}.
Here the structure of the groups admitting such type of change of representation, a characterization of the induced semigroups and a prototype for the quasi-affinities are given in the following terms:
a unitary group $\{U_t\}$ and a contraction semigroup $\{W_t\}$ satisfy the intertwining relation (\ref{Int.Rel}) for a quasi-affinity $\LL$ if and only if (iff) $\{W_t\}$ belongs to the class $C_{\cdot 1}$ and 
$\{U_t\}$ is unitarily equivalent to the residual group $\{R_t\}$ of the minimal isometric dilation of $\{W_t\}$;
in such situation the quasi-affinity intertwining $\{R_t\}$ and $\{W_t\}$ is explicitly given.
In other words, a contraction semigroup has unitary quasi-affine transforms iff it is in the class $C_{\cdot 1}$, and there exist universal representatives for the unitary quasi-affine transforms: the residual groups, their functional models or, equivalently, the unitary $*$-asymptotic groups given by K\'erchy \cite{KER87}. 
Clear advantages derive from the existence of such universal representatives, as for example  determining spectral properties or clarifying the links between intertwining quasi-affinities $\LL$ and time operators $T$ \cite{G2}.

The paper is organized as follows.
Section \ref{sc2} includes the main results: Subsection \ref{s4} deals with single operators and Subsection \ref{sschs} with groups and semigroups. Proofs and additional comments are collected in Section \ref{s2}. Some of the consequences of these results are given in Sections \ref{sfm}, \ref{s5} and \ref{sirsm}. The functional models are described in Section \ref{sfm}, some spectral properties and relations derived in Section \ref{s5} and, by way of conclusion,  the results are applied to MPC theory in the context of statistical mechanics  in Section \ref{sirsm}.  
For the sake of completeness, the work ends with an Appendix about similarity relation.


\section{Main results}\label{sc2}

Let us begin by recalling some definitions:

\begin{definition}\label{dqss}\rm
a) Let $\H_1$ and $\H_2$ be Hilbert spaces. Let $\L(\H_1,\H_2)$ denote the space of bounded linear operators from $\H_1$ into $\H_2$. We shall write $\L(\H_1)=\L(\H_1,\H_1)$. Given operators $W_1\in\L(\H_1)$ and $W_2\in\L(\H_2)$, the {\it intertwining set} $\I(W_1,W_2)$ is given by
$$
{\mathcal I}(W_1,W_2):=\{\LL\in \L(\H_1,\H_2):\LL W_1=W_2\LL\}\,.
$$

b) By a {\it quasi-affinity} from $\H_1$ to $\H_2$ we mean a linear, one-to-one and continuous transformation $\LL $ from $\H_1$ onto a dense subspace in $\H_2$, so that $\LL ^{-1}$ exists on this dense domain, but is not necessarily continuous. 
For bounded operators $W_1$ acting on $\H_1$ and $W_2$ on $\H_2$ we say that $W_1$ is a {\it quasi-affine transform} of $W_2$ if there exists an quasi-affinity $\LL\in\I(W_1,W_2)$. 
$W_1$ and $W_2$ are called {\it quasi-similar} if they are quasi-affine transforms of one another.

c) By an {\it affinity} from $\H_1$ to $\H_2$ we mean a linear, one-to-one and bicontinuous transformation $\LL:\H_1\to\H_2$. 
Operators $W_1$ and $W_2$ are called {\it similar} if there exists an affinity $\LL\in\I(W_1,W_2)$ (and consequently $\LL^{-1}\in\I(W_2,W_1)$). Similarity is an equivalence relation.
$W_1$ and $W_2$ are called {\it unitarily equivalent} if $\I(W_1,W_2)$ contains a unitary operator.
\end{definition}

While similarity is a rather strong relation, which preserves, for example, the spectrum, quasi-similarity does not have such strong implications.
Conditions for similarity of contractions and unitary operators are well known and some of them collected in Proposition \ref{NF.C.IX.1.4} of Appendix \ref{as1}. Here we are interested in the more general situation MPC theory deals with: the study of unitary quasi-affine transforms of contractions.

\subsection{Unitary quasi-affine transforms of contractions}\label{s4}

Let $\H\subset\K$ be two Hilbert spaces. Following the terminology and notations used by Sz.-Nagy and Foia\c s in \cite{NAGY-FOIAS}, for operators $A:\H\to\H$ and $B:\K\to\K$ we write 
$A=\text{pr}\, B$
when $(Ah,h')=(Bh,h')$ for all $h,h'\in\H$ or, equivalently, $Ah=P_\H Bh$ for all $h\in\H$, where $P_\H$ denotes the orthogonal projection of $\K$ onto $\H$.
We call $B$ a {\it dilation} of $A$ if 
\b\label{rdila}
A^n=\text{pr }B^n,\quad n=1,2,\ldots
\e
Two dilations of $A$, say $B$ on $\K$ and $B'$ on $\K'$, are called {\it isomorphic} if there exists a unitary operator $U:\K\to\K'$ such that $Uh=h$ for $h\in\H$ and $B'=U^{-1}BU$.
For every {\it contraction} $W$ on a Hilbert space $\H$ there exist an isometric dilation $U_+$ on some Hilbert space $\K_+\supset\H$ and a unitary dilation $U$ on some Hilbert space $\K\supset\H$, which are moreover minimal in the sense that
$\K_+=\bigvee_0^\infty U_+^n\H$ and $\K=\bigvee_{-\infty}^\infty U^n\H$.
These minimal isometric and unitary dilations are determined up to isomorphism, {\it c.f.} \cite[Section I.4]{NAGY-FOIAS}.
In what follows we  consider the minimal isometric dilation $U_+$ of $W$ embedded in its minimal unitary dilation $U$ in the following way:
$\K_+:=\bigvee_{0}^\infty U^n\H$ and $U_+:=U_{|\K_+}$,
i.e., $U_+$ is the restriction to $\K_+$ of $U$.

The isometric minimal dilation $U_+$ on $\K_+$ admits a unique {\it Wold decomposition} \cite[Th.I.1.1]{NAGY-FOIAS} into a unitary part and a unilateral shift (see Lemma \ref{NF.T.I.3.2} below).
The unitary part $R,\RR$ is given by 
\b\label{NFW}
\RR:=\bigcap_{n=0}^\infty U_+^n \H\,,\quad R:=U_+|\RR\,,
\e
and is called the {\it residual part} of $U_+,\K_+$.

From now on
$$
P_{\mc M}
$$ 
will always denote the orthogonal projection from $\K_+$ or $\K$ onto a closed subspace $\mc M$. Which space, $\K_+$ or $\K$, will be clear by the context.

One has the following characterization of contractions having unitary quasi-affine transforms:

\begin{proposition}\label{CN11.3}
Let $\H$ be a Hilbert space and let $W$ be a contraction on $\H$ such that $\text{Ker}\,W=\{0\}$.
The following statements are equivalent:

(i) $W$ has unitary quasi-affine transforms;

(ii) $W$ belongs to the class $C_{\cdot 1}$, i.e., 
\b\label{cc.1}
\lim_{n\to\infty} W^{*n}h\neq 0 \text{ for each non-zero }h\in \H\,. 
\e

In such case the residual part $R$ of the minimal isometric dilation of $W$ is a (unitary) quasi-affine transform of $W$ and 
\b\label{intop}
\LL_0:=P_\H|\RR=(P_\RR|\H)^*
\e
is an intertwining quasi-affinity belonging to $\I(R,W)$.
\end{proposition}

Actually, every unitary quasi-affine transform is unitarily equivalent to the corresponding residual part:

\begin{proposition}\label{CN11.1}
Let $\H$ be a Hilbert space and let $W$ be a contraction on $\H$ such that $\text{Ker}\,W=\{0\}$.
Then every unitary quasi-affine transform of $W$ is unitarily equivalent to the residual part of the minimal isometric dilation of $W$.
\end{proposition}

The concepts of {\it isometric and unitary asymptotes}  were introduced by 
K\'erchy \cite{KER89} for power bounded operators. Here for definitions we adopt universal properties and restrict attention to contractions and their unitary $*$-asymptotes. 

\begin{definition}\label{uaa}\rm
Let $\H_1$ and $\H_2$ be Hilbert spaces, let $W\in\L(\H_1)$ be a contraction, $U\in\L(\H_2)$ a unitary operator and $\LL\in {\mathcal I}(U,W)$. We 
The pair $(\LL,U)$ is called a {\it unitary $*$-asymptote} of $W$ if it has the following {\it universal property}:
For any unitary operator $V$ and any intertwining operator $\Theta\in{\mathcal I}(V,W)$ there exists a unique operator $\Theta_*^{(a)}\in{\mathcal I}(V,U)$ such that $\Theta=\LL\Theta_*^{(a)}$.
\end{definition}

Now consider each unitary quasi-affine transform $U$ of $W$ together with the corresponding intertwining quasi-affinity $\LL\in\I(U,W)$ in a pair $(\LL,U)$. Then the unitary $*$-asymptotes of $W$ are just its unitary quasi-affine transforms:

\begin{corollary}\label{coro1}
Let $\H$ be a Hilbert space and let $W$ be a contraction on $\H$ of class $C_{\cdot1}$ and such that $\text{Ker}\,W=\{0\}$.
Then the set of unitary $*$-asymptotes of $W$ and the set of unitary quasi-affine transforms of $W$ coincide. Both sets coincide with the set of unitary operators which are unitarily equivalent to the residual part of the minimal isometric dilation of $W$.
\end{corollary}

Concrete realizations of unitary $*$-asymptotes have been given by K\'erchy \cite{KER89} in terms of Banach limits (see Remark \ref{ruaa} below).


\subsection{Intertwining unitary groups and contraction semigroups}\label{sschs}

Due to relation (\ref{rdila}) for dilations, 
the results of Section \ref{s4} for a single contraction $W$ and its unitary quasi-affine transforms $(\LL,U)$ extend to the corresponding discrete semigroup $\{W^n\}_{n\in\N}$ and group $\{U^n\}_{n\in\Z}$ in a natural way:  $\LL\in\I(U^n,W^n)$ and $U^n$ is unitarily equivalent to the residual part $R^n$ for every $n\in\N$. In order to study intertwining relations between unitary groups and contraction semigroups for continuous time parameter, we will utilize their cogenerators and  the Sz.-Nagy and Foia\c s functional calculus (see Section \ref{psschs} for details).

By a {\it (continuous one-parameter) semigroup} on a Hilbert space $\H$ we mean a family $\{W_t\}_{t\geq 0}\subset \L(\H)$ with the following properties:
(1) $W_tW_s=W_{t+s}$, for $t,s\geq 0$;
(2) $W_0=I$;
(3) $\slim_{t\to s} W_th=W_{s}h$, for each $s\geq 0$ and $h\in \H$, i.e. 
$\slim_{t\to s} W_t=W_{s}$, where $\slim$ denotes limit in strong sense in both $\H$ and $\L(\H)$.
A family $\{W_t\}_{t\in\R}$ is called a {\it (continuous one-parameter) group} if it satisfies (2) as well (2) and (3) for $t,s\in\R$. Thus, from (1) and (2), $W_{-t}=W_t^{-1}$.

In what follows $\{W_t\}_{t\geq 0}$ shall denote a semigroup of contractions on a Hilbert space $\H$
and $\{U'_t\}_{t\in\R}$ a unitary group on a Hilbert space $\H'$.

\begin{definition}\rm
A unitary group $\{U'_t\}_{t\in\R}$ is called a {\it unitary quasi-affine transform} of a contraction semigroup $\{W_t\}_{t\geq 0}$ if there exists a quasi-affinity $\LL\in \I(U'_t,W_t)$ for $t\geq0$.
Two unitary groups $\{U'_t\}_{t\in\R}$ and $\{U''_t\}_{t\in\R}$ are called {\it unitarily equivalent}  
if there exists a unitary operator $V\in \I(U'_t,U''_t)$ for every $t\in\R$.
\end{definition}

For a contraction semigroup $\{W_t\}_{t\geq 0}$ with infinitesimal generator $A$, ($W_t=\exp(tA)$), the {\it cogenerator} $W$ of $\{W_t\}_{t\geq 0}$ is the Cayley transform of $A$ given by
$$
W=(A+I)(A-I)^{-1},\quad A=(W+I)(W-I)^{-1}.
$$
The cogenerator $W$ is a contraction which does not have $1$ among its eigenvalues.
Moreover, the semigroup $\{W_t\}_{t\in\R^+}$ consists of normal, selfadjoint, isometric or unitary operators iff its cogenerator $W$ is normal, selfadjoint, isometric or unitary, respectively. 
Moreover, the residual group $\{R_t\}_{t\in\R}$ of a contraction semigroup $\{W_t\}_{t\in\R^+}$ (i.e., $R_t$ is the residual part corresponding to $W_t$ for every $t\geq0$) is just the unitary group on $\RR$ whose cogenerator is the residual part $R,\RR$ of for the cogenerator $W$ of $\{W_t\}_{t\in\R^+}$. 
See \cite[Sect.III.8-9]{NAGY-FOIAS}.

Here we have the extended versions of Propositions \ref{CN11.3} and \ref{CN11.1} for groups and semigroups. 

\begin{theorem}\label{TN11.3}
Let $\H$ be a Hilbert space and let $\{W_t\}_{t\geq 0}$ be a contraction semigroup on $\H$ with cogenerator $W$  such that $\text{Ker}\,W=\{0\}$. The following statements are equivalent:

i) there exist unitary quasi-affine transforms $\{U'_t\}_{t\in\R}$ of $\{W_t\}_{t\geq 0}$;

ii) $\{W_t\}_{t\geq 0}\in C_{\cdot 1}$, i.e.
\b\label{gc.1}
\lim_{t\to\infty} W^*_th\neq 0 \text{ for each non-zero }h\in \H 
\e

In such case the group of residual parts $\{R_t\}_{t\in\R}$ for  $\{W_t\}_{t\geq 0}$ is a unitary quasi-affine transform of $\{W_t\}_{t\geq 0}$ and the quasi-affinity $\LL_0$ defined in 
(\ref{intop}) belongs to $\I(R_t,W_t)$ for every $t\geq0$.
\end{theorem}

\begin{theorem}\label{TN11.1}
Let $\H$ be a Hilbert space and let $\{W_t\}_{t\geq 0}$ be a contraction semigroup on $\H$ of class $C_{\cdot,1}$ and with cogenerator $W$  such that $\text{Ker}\,W=\{0\}$. Then every unitary quasi-affine transform $\{U'_t\}_{t\in\R}$ of $\{W_t\}_{t\geq 0}$ is unitarily equivalent to the group of residual parts $\{R_t\}_{t\in\R}$ for $\{W_t\}_{t\geq 0}$.
\end{theorem}

\begin{remark}\rm \label{rcd}
The conditions given above in terms of $\{W_t\}_{t\geq 0}$ and $\{U'_t\}_{t\in\R}$ can be written in terms of their respective cogenerators $W$ and $U'$. Actually, $\LL\in\I(U',W)$ is equivalent to $\LL\in\I(U'_t,W_t)$ for $t\geq0$ (see Lemma \ref{Np.0} below), and conditions (\ref{cc.1}) and (\ref{gc.1}) are also  equivalent (see \cite[Sect.III.9]{NAGY-FOIAS} for details).
\end{remark}

Corollary \ref{coro1} extends in a similar way.


\section{Proofs and additional remarks}\label{s2}

Let us pass to prove the results of Section \ref{sc2}.
Some of their consequences will be given afterwards.

\subsection{Intertwining unitary and contraction operators}\label{ps4}

Recall that for an isometry $V$ on a Hilbert space $\H$ 
a subspace $\L\subset\H$ is called {\it wandering} if $V^n\L\perp V^m\L$ for every pair of integers $m,n\geq 0$, $m\neq n$ --actually, since $V$ is an isometry, it suffices that $V^n\L\perp\L$ for $n\in\N$--.
The orthogonal sum
$$
M_+(\L):=\bigoplus_{n=0}^\infty V^n\L
$$
satisfies $VM_+(\L)=\oplus_1^\infty V^n\L=M_+(\L)\ominus\L$.
If $U$ is a unitary operator on $\H$ and $\L$ is a wandering subspace for $U$, since $U^{-1}$ is also isometric, $U^m\L\perp U^n\L$ for all integers $m\neq n$. The subspace
$$
M(\L):=\bigoplus_{n=-\infty}^\infty U^n\L
$$
reduces $U$. $M(\L)$ does not determine $\L$, only its dimension.

For a contraction $W$ on the Hilbert space $\H$ with minimal unitary dilation $U$ on $\K$ the subspaces 
$\L:=\overline{(U-W)\H}$ and $\L^*:=\overline{(U^*-W^*)\H}$ (the overbar denotes adherence)
are wandering subspaces for $U$ and the space $\K$ can be decomposed into the orthogonal sum
$$
\K=\cdots\oplus U^{*2}\L^*\oplus U^*\L^*\oplus\L^*\oplus\H\oplus \L\oplus U\L\oplus U^2\L\oplus\cdots
$$
$M(\L)$ and $M(\L^*)$ reduce $U$ and hence the same is true for the subspaces
$\RR:=\K\ominus M(\L^*)$ and $\RR_*:=\K\ominus M(\L)$.
The residual part and dual residual part of $U$ are the unitary operators
$R:=U_{|\RR}$ and $R_*:=U_{|\RR_*}$. 
Now consider the subspace
$\L_*:=U\L^*=\overline{(I-UW^*)\H}$.
Then $\L$ and $\L_*$ are wandering subspaces for the minimal isometric dilation $U_+$ of $W$ (and hence for $U$) such that $\L\cap\L_*=\{0\}$ and
$$
\K_+=\H\oplus M_+(\L)=\RR\oplus M_+(\L_*).
$$

The following result is a lifting theorem for operators $\LL $ intertwining contractions and unitary operators. In this case an explicit expression (\ref{n.e}) for the lifting $\LL _+$ is given and the relevant part of the dilation is the residual one.
Expressions similar to (\ref{n.e}) have been considered in the study of Pt\'ak ge\-ne\-ra\-li\-za\-tion of Toeplitz and Hankel operators, see \cite{Pt97} and references therein. Here we give a proof based on that of the classical lifting theorem \cite[Th.II.2.3]{NAGY-FOIAS}.

\begin{lemma}\label{N1}
Let $\H$ and $\H'$ be Hilbert spaces, let $W$ be a contraction on $\H$ with minimal isometric dilation $U_+$ on $\K_+$, and let $U'$ be a unitary operator on $\H'$.
For every bounded operator $\LL :\H'\to\H$ satisfying the intertwining relation
\b\label{n.a}
W\LL =\LL  U'
\e
the unique bounded operator $\LL _+:\H'\to\K_+$ satisfying 
\b\label{n.b}
U_+\LL _+=\LL _+U'\,,
\e
\b\label{n.d}
||\LL ||=||\LL _+||\,.
\e
\b\label{n.c}
\LL =P_\H \LL _+\,,
\e
is of the form
\b\label{n.e}
\LL _+=\slim_{n\to\infty} U_+^n\LL  U'^{-n}\,.
\e
Moreover, 
the range of $\LL _+$ is contained in the residual part 
$\RR$ of $\K_+$, i.e.
\b\label{P1}
\LL _+ \H'\subseteq \RR\,.
\e
\end{lemma}

\begin{proof}
Since $\K_+=\H\oplus M_+(\L)$ the general form of an operator $\LL _+:\H'\to\K_+$ satisfying (\ref{n.c}) is
\b\label{2.14}
\LL _+=\LL +B_0+U_+B_1+U_+^2B_2+\cdots,
\e
where each $B_n$ is an operator from $\H'$ into $\L$.
From (\ref{2.14}) we deduce
$$
U_+\LL _+-\LL _+U'=\sum_{n=0}^\infty U_+^n(B_{n-1}-B_nU'),
$$
with $B_{-1}=U_+\LL -\LL  U'$.
Because of (\ref{n.a}) we have $B_{-1}=(U_+-W)\LL $ and thus $B_{-1}$ is an operator from $\H'$ into $\L$.
Being $U'$ unitary, in order that $\LL _+$ satisfies (\ref{n.b}), it is therefore necessary and sufficient that
$$
B_n=B_{n-1} U'^{-1} \text{ for }n=0,1,\ldots,\quad B_{-1}=(U_+-W)\LL ,
$$
so that
$$
B_n=(U_+-W)\LL  U'^{-(n+1)} \text{ for }n=0,1,\ldots
$$
and, using (\ref{n.a}),
$$
\begin{array}{rl}
\LL _+ \!\!\! & \ds = \LL  +\sum_{n=0}^\infty U_+^n(U_+-W)\LL  U'^{-(n+1)}
\\[2ex]
& \ds = \LL  +\sum_{n=0}^\infty U_+^{n+1}\LL  U'^{-(n+1)}-U_+^nW\LL  U'^{-(n+1)}
\\[2ex]
& \ds = \LL  +\sum_{n=0}^\infty U_+^{n+1}\LL  U'^{-(n+1)}-U_+^nW\LL  U'^{-(n+1)}
\\[2ex]
& \ds = \LL  +\sum_{n=0}^\infty U_+^{n+1}\LL  U'^{-(n+1)}-U_+^n\LL  U'^{-n}
\\[2ex]
& \ds = \LL  -\LL +\slim_{n\to\infty} U_+^n\LL  U'^{-n}=\slim_{n\to\infty} U_+^n\LL  U'^{-n},
\end{array}
$$
being the last limit in strong sense on $L(\H)$ because we are dealing with an orthogonal sum and for the $N$-th sum and each $h'\in\H'$, since $U_+$ is an isometric extension of $W$, we have  
$$
\begin{array}{l}
\ds ||(\LL  +\sum_{n=0}^{N-1} U_+^nB_n)h'||^2
= ||\LL  h'||^2 +\sum_{n=0}^{N-1} ||B_nh'||^2
\\[2ex]
\hspace{10ex} \ds = ||\LL  h'||^2 +\sum_{n=0}^{N-1} || (U_+-W)\LL  U'^{-(n+1)}h'||^2
\\[2ex]
\hspace{10ex} \ds = ||\LL  h'||^2 +\sum_{n=0}^{N-1} \left(|| \LL  U'^{-(n+1)}h'||^2 -||\LL  U'^{-n}h'||^2\right)
\\[2ex]
\hspace{10ex} \ds = ||\LL  h'||^2 -||\LL  h'||^2+ || \LL  U'^{-N}h'||^2
\\[2ex]
\hspace{10ex} \ds \leq ||\LL ||^2\,||U'^{-N}h'||^2= ||\LL ||^2\,||h'||^2.
\end{array}
$$
Moreover, since for all $\LL _+$ satisfying (\ref{n.c}) the inequality $||\LL ||\leq||\LL _+||$ holds, 
we have $||\LL ||=||\LL _+||$.

Now, recall that $\K_+= M_+(\L_*)\oplus\RR$ corresponds to the Wold decomposition of $U_+$, being ${U_+}_{|\RR}$ unitary and ${U_+}_{|M_+(\L_*)}$ a unilateral shift. Thus, being $U'$ unitary, from (\ref{NFW}) and (\ref{n.b}), we have
$$
\LL _+\H'=\bigcap_{n=0}^\infty \LL _+ U'^n \H' = \bigcap_{n=0}^\infty U_+^n \LL _+ \H'\subseteq
\bigcap_{n=0}^\infty U_+^n \K_+=\RR\,,
$$
so that (\ref{P1}) is proved.
\end{proof}

\begin{remark}\label{RN1}\rm  
The operator $\LL _+:\H'\to\K_+$ of Lemma \ref{N1} can also be considered as an operator from $\H'$ into the space $\K\supseteq\K_+$ where the minimal unitary dilation $U$ of $W$ is defined. We will denote this operator by $\LL _+$ as well. Obviously $\LL _+:\H'\to\K$ is of the form 
\b\label{n2.e}
\LL _+=\slim_{n\to\infty} U^n\LL  U'^{-n}
\e
and satisfies the conditions
$U\LL _+=\LL _+U'$,
$||\LL ||=||\LL _+||$ and
$\LL =P_\H \LL _+$.
From now on we shall use either meanings of $\LL _+$ without causing confusion.
\end{remark}


In order to prove Propositions \ref{CN11.1} and \ref{CN11.3} we will need the following technical Lemmas.

\begin{lemma}\label{P5}
Let $\H$ be Hilbert space and let $W$ be a contraction on $\H$ with minimal isometric and unitary dilations $U_+$ and $U$ on $\K_+$ and $\K$, respectively, and $R$, $\RR$ the corresponding residual part. For a non-zero $h\in\H$ the following assertions are equivalent:
\begin{itemize}
\item[(a)]
$h\perp P_\H\RR$;
\item[(b)]
$\ds \slim_{n\to\infty} {W^*}^n h=0$.
\end{itemize}
\end{lemma}
 
\begin{proof}
Let $h$ be a non-zero vector such that $h\in\H$ and $h\perp P_\H\RR$ or, equivalently, such that $h\in \H$ and $h\perp\RR$.
Since $K_+=\RR\oplus M_+(\L_*)$, $h\in M_+(\L_*)$ and hence $h$ has an orthogonal expansion $h=\sum_{m=0}^\infty U^mh_m$, where $h_m\in \L_*$ and $||h||^2=\sum_{m=0}^\infty ||h_m||^2$.
Moreover, since $W^*=\text{pr}\,U^{-1}$ and $U^{-\nu}\L_*\perp\H$ for $\nu>0$, we have
$$
{W^*}^nh=P_\H U^{-n}h=P_\H \sum_{m=0}^\infty U^{m-n}h_m
=P_\H \sum_{m\geq n} U^{m-n}h_m,
$$
so that
$\ds\slim_{n\to\infty} {W^*}^nh=\slim_{n\to\infty} \sum_{m\geq n} U^{m-n}h_m=0$
and (a)$\Rightarrow$(b) is proved. 
Now, to prove (b)$\Rightarrow$(a) assume that for a non-zero $h\in\H$ one has $\ds\slim_{n\to\infty} {W^*}^n h=0$. Then
$$
\sum_{k=0}^{n-1} U^{k+1}(U^*-W^*){W^*}^k h = h-U^n{W^*}^nh \in M_+(\L_*)
$$
and $\ds\slim_{n\to\infty} (h-U^n{W^*}^nh) =h\in M_+(\L_*)$. Thus, $h\perp P_\H\RR$.
\end{proof}

In what follows we come back to the Hilbert spaces $\H$ and $\H'$ of the lifting lemma \ref{N1}.

\begin{lemma}\label{N11}
Let $\H$ and $\H'$ be Hilbert spaces. Let $W$ be a contraction on $\H$ with $\text{Ker}\,W=\{0\}$, minimal isometric and unitary dilations $U_+$ and $U$ on $\K_+$ and $\K$, respectively, and residual part $R$, $\RR$. Let $U'$ be a unitary operator on $\H'$ which is a quasi-affine transform of $W$, i.e. there exists a quasi-affinity $\LL :\H'\to\H$ such that
\b\label{n11.1}
W\LL =\LL  U',
\e
and let $\LL _+:\H'\to\K_+$ (or $\LL _+:\H'\to\K$) be the lifting operator 
given in (\ref{n.e}) or (\ref{n2.e}). Then:
\begin{itemize}
\item[(a)]
$W\in C_{\cdot 1}$, i.e. $W^{*n}h$ does not converge to $0$ for each non-zero $h\in \H$;
\item[(b)]
$\RR\cap M_+(\L)=\{0\}$;
\item[(c)]
$\LL _+$ is a quasi-affinity from $\H'$ into $\RR$ such that
$R\LL _+=\LL _+ U'$.
\end{itemize}
\end{lemma}

\begin{proof}
(a) Since $\LL $ is a quasi-affinity from $\H'$ to $\H$, we have $\H=\overline{\LL \H'}$. Property (\ref{n.c}) says that $\LL =P_\H \LL _+$. By (\ref{P1}), $\LL _+\H'\subseteq\RR$. Thus,
$$
\H= \overline{\LL \H'}=\overline{P_\H \LL _+\H'}\subseteq \overline{P_\H\RR}\subseteq\H
$$
and therefore $\overline{P_\H\RR}=\H$. But, by Lemma \ref{P5}, $\overline{P_\H\RR}=\H$ implies that $W^{*n}h$ does not converge to $0$ for each non-zero $h\in \H$.

(b) Suppose there exists a non-zero $k\in \RR\cap\H^\perp$. Then $k\perp M(\L^*)$ and $k\perp\H$, so that $k\in M_+(\L)$ and hence $k$ has an orthogonal expansion $k=\sum_{n=0}^\infty U^nk_n$, where $k_n\in\L$ and $||k||^2=\sum_{n=0}^\infty ||k_n||^2$.
Since $k\neq 0$, there is at least one non-zero $k_n$; let $k_\nu$ be the first of these non-zero terms.
Then we have
\b\label{II.3.5}
U^{-\nu-1}k=U^{-1}k_\nu+\sum_{\mu=0}^\infty U^\mu k_{\nu+\mu+1}.
\e
Since $k\in\RR$ and $\RR$ reduces $U$, also $U^{-\nu-1}k$ belongs to $\RR=\K\ominus M(\L^*)$ and, in particular, $U^{-\nu-1}k\perp \L^*$. Moreover, $U^\mu\L\perp\L^*$ for $\mu\geq 0$ and we deduce from 
(\ref{II.3.5}) that $U^{-1}k_\nu\perp\L^*$ and $k_\nu\perp U\L^*$.
Since $\H\oplus\L=U\L^*\oplus U\H$, we conclude that $k_\nu\in U\H$. Thus there exists an $h\in\H$ such that $k_\nu=Uh$; consequently $P_\H k_\nu=P_\H Uh=Wh$. Since $\L\perp\H$, we have $P_\H k_\nu=0$ and hence $Wh=0$. But $k_\nu\neq0$ implies $h\neq0$, and this is in contradiction with $\text{Ker}\,W=\{0\}$. 

(c) By (\ref{P1}), $\LL _+\H'\subseteq\RR$. We must prove that $\LL _+$ is injective and $\overline{\LL _+\H'}=\RR$. The injectivity of $\LL _+$ follows from that of $\LL $. Indeed, if there exist $h'_1,h'_2\in\H'$ such that $\LL _+h'_1=\LL _+h'_2$, then $P_\H \LL _+h'_1=P_\H \LL _+h'_2$, that is $\LL  h'_1=\LL  h'_2$ and thus $h'_1=h'_2$. 
Now suppose that $\overline{\LL _+\H'}\neq\RR$, i.e. that there exists a non-zero $k\in\RR$ such that $k\perp \LL _+\H'$ and, then,
$(k,\LL _+h')=0$ for all $h'\in\H'$.
Taking into account expression (\ref{n.e}) for $\LL _+$ and the relation $\LL =P_\H \LL _+$ (see Lemma \ref{N1} and Remark \ref{RN1}) we have then
$$
\begin{array}{rl}
(k,\LL _+h')& \ds =\lim_{n\to\infty} (k,U^n\LL  U'^{-n}h')
 =\lim_{n\to\infty} (U^{-n}k,P_\H \LL _+ U'^{-n}h')=
\\[2ex]
&\ds =\lim_{n\to\infty} (U'^{n}\LL _+^*P_\H U^{-n}k,h')=0,\quad \text{ for all }h'\in\H'.
\end{array}
$$
But this is equivalent to 
$\slim_{n\to\infty} U'^{n}\LL _+^*P_\H U^{-n}k=0$,
which, since $U'$ is unitary, coincides with
$\slim_{n\to\infty} \LL _+^*P_\H U^{-n}k=0$.
Thus,
$$
\begin{array}{l}
\ds \lim_{n\to\infty} (\LL _+^*P_\H U^{-n}k,h')=
\lim_{n\to\infty} (P_\H U^{-n}k,P_\H \LL _+h')=
\\[2ex]
\ds =\lim_{n\to\infty} (P_\H U^{-n}k,\LL  h')=0,\quad \text{ for all }h'\in\H',
\end{array}
$$
and, since $\LL $ is quasi-affinity, $\LL \H'$ is dense in $\H$ and this implies
\b\label{xx1}
\slim_{n\to\infty} P_\H U^{-n}k=0.
\e
Now recall that $\K_+=\H\oplus M_+(\L)=\RR\oplus M_+(\L_*)$ and that $\RR$ reduces $U$ to its residual part $R$ and then $U^{-n}k=R^{-n}k\in\RR$ for all $k\in\RR$ and $n\in\Z$. Therefore (\ref{xx1}) implies 
$\slim_{n\to\infty} U^{-n}k\in\RR\cap M_+(\L)$, but, from (b), $\RR\cap M_+(\L)=\{0\}$ and we have
$\slim_{n\to\infty} U^{-n}k=0$, 
only possible if $k=0$, since $U$ is unitary. This proves  $\overline{\LL _+\H'}=\RR$.
\end{proof} 

\begin{remark} \rm
Under the conditions of Lemma \ref{N11} except Ker$\,W=\{0\}$, we have $\text{Ker}\,W \cap\text{Rang}\,\LL =\{0\}$. Indeed, if there exists a non-zero $h\in \text{Ker}\,W$ and $h=\LL  h'$ for some $h'\in\H'$, from (\ref{n11.1}),
$0=Wh=W\LL  h'=\LL  U'h'$,
so that $U'h'\in\text{Ker}\,\LL $ and therefore $\LL $ cannot be a quasi-affinity from $\H'$ into $\H$.
\end{remark}

The main assertion of Lemma \ref{N11} is that, given a unitary quasi-affine transform $U'$ of a contraction $W$ (with $\text{Ker}\,W=\{0\}$) and  a quasi-affinity $\LL$ intertwining both operators, $U'$ is also a quasi-affine transform of the residual part $R$ of the minimal isometric dilation of $W$ and the lifting $\LL _+$ of $\LL $ is a quasi-affinity intertwining $U'$ and $R$.
An immediate corollary of this fact is that then $U'$ and $R$ are unitarily equivalent.

\begin{proof}[of Proposition \ref{CN11.1}]
This follows from Lemma \ref{N11}.(c) and the fact that, 
if a unitary operator $U_1$ on $\H_1$ is a quasi-affine transform of a unitary operator $U_2$ on $\H_2$, then $U_1$ and $U_2$ are unitarily equivalent \cite[Prop.II.3.4]{NAGY-FOIAS}.
\end{proof}

\begin{remark}\rm
Under the conditions of Proposition \ref{CN11.1}, let $U'$ be a quasi-affine transform of $W$ and $\LL\in\I(U',W)$ an intertwining quasi-affinity. In the light of Lemma \ref{N11}.(c), the unitary operator $V\in\I(U',R)$ performing the unitary equivalence of $U'$ and the residual part $R$ can be written in terms of the lifting $\LL_+$ given in Lemma \ref{N1}. Indeed,  $\LL_+|\LL_+|^{-1}$ extends by continuity to $V$, where $|\LL_+|=(\LL_+^*\LL_+)^{1/2}$.
\end{remark}

Other consequence of Lemma \ref{N11} is the characterization of contractions having unitary quasi-affine transforms given in Proposition \ref{CN11.3}:

\begin{proof}[of Proposition \ref{CN11.3}]
First part follows from Lemma \ref{N11}.(a). Second part is then a consequence of \cite[Prop.II.3.5]{NAGY-FOIAS}.
\end{proof} 

\begin{proof}[of Corollary \ref{coro1}]
Theorem 2 and subsequent comments in \cite{KER89} plus Propositions \ref{CN11.3} and \ref{CN11.1} lead to the result.
\end{proof}

\begin{remark}\label{ruaa}\rm
K\'erchy \cite{KER89} dealt with the following concrete realizations of the asymptotes of a power bounded operator: Let $L$ denote a Banach limit, that is, a positive linear functional on the sequence space $l^\infty(\N)$ with the properties $L(1,1,\ldots)=1$ and $L(c_1,c_2,c_3,\ldots)=L(c_2,c_3,\ldots)$. 
Let $\H$ be a Hilbert space with inner product $(\cdot,\cdot)$ and let $W$ be a power bounded operator on $\H$. Setting
$$
[h,g]:=L\big(\{(T^nh,T^ng)\}_n\big)\,,\quad (h,g\in\H)\,,
$$
and $\H_0:=\{h\in\H:[x,x]=0\}$, the quotient $\H/\H_0$ is an inner product space with the inner product $[h+\H_0,g+\H_0]:=[h,g]$. Let $\H_+^{(a)}=\H_{+,W}^{(a)}$ denote the resulting Hilbert space obtained by completion.
Since $L$ is a Banach limit, $[Th,Tg]=[h,g]$ for every $h,g\in\H$. Hence $W_0:h+\H_0\mapsto Wh+\H_0$ is a well-defined isometry on $\H/\H_0$.
Let $W_+^{(a)}$ denote the continuous extension of $W_0$ to the space $\H_+^{(a)}$.
According K\'erchy, the isometry $W_+^{(a)}$ is the {\it isometric asymptote} of $W$, the minimal unitary dilation of $W_+^{(a)}$, denoted by $W^{(a)}$ and acting on the space $\H^{(a)}=\H_{W}^{(a)}$, is the {\it unitary asymptote} of $W$, and the operator $W_*^{(a)}:=\big((W^*)^{(a)}\big)^*$ acting on $\H_{*,W}^{(a)}=\H_{W^*}^{(a)}$ is the {\it unitary $*$-asymptote} of $W$.
Associated with them, the following intertwining operators:
the quotient map $\LL_+^{(a)}=\LL_{+,W}^{(a)}$ from $\H$ into $\H_+^{(a)}$: $\LL_+^{(a)}:h\mapsto h+\H_0$, the corresponding embedding $\LL^{(a)}=\LL_{W}^{(a)}$ of $\H$ into $\H^{(a)}$,  and $\LL_*^{(a)}:=(\LL_{W^*}^{(a)})^*$. Clearly, $\LL_+^{(a)}\in{\mathcal I}(W,W_+^{(a)})$, $\LL^{(a)}\in {\mathcal I}(W,W^{(a)})$ and $\LL_+^{(a)}h=\LL^{(a)}h$ for every $h\in\H$, and $\LL_*^{(a)}\in{\mathcal I}(W_*^{(a)},W)$.
\end{remark}


\subsection{Intertwining unitary groups and contraction semigroups}\label{psschs}

In order to extend the results of Section \ref{s4} for single operators to groups and semigroup of operators let us begin by introducing some basic concepts about the functional calculus for contractions on Hilbert spaces given by Sz.-Nagy and Foia\c s \cite[Chapter III]{NAGY-FOIAS}. Let $A$ be the algebra of functions holomorphic in the open unit disc $D$ and continuous on $\overline D$ given by
$$
A:=\Big\{a(\l)=\sum_{k=0}^\infty c_k\l^k \text { with } \sum_{k=0}^\infty |c_k|<\infty \Big\}\,,
$$
with involution $a(\l)\to \tilde a(\l)=a^*(\l^*)$.
Given a contraction $W$ on a Hilbert space $\H$ and $a=\sum c_k\l^k\in A$ we can define
\b\label{at}
a(W):=\sum_{k=0}^\infty c_k W^k,\quad a(W)^*=\tilde a(W^*)\,,
\e
the series converging in operator norm. If $W$ is a normal operator with spectral representation $W^n=\int_{\sigma(W)} \l^n\,dK$ ($n=0,1,\ldots$), definition (\ref{at}) coincides with the usual one $a(W)= \int_{\sigma(W)} a(\l)\,dK$, since $\sigma(W)\subset \overline D$ and $\sum c_k\l^k$ converges uniformly on $\overline D$. 
For every function $u$ holomorphic in $D$ the function $u_r(\l):=u(r\l)$ ($0<r<1$) belongs to $A$.
Denote by $H_W^\infty$ the set of functions $u$ in the Hardy class $H^\infty$ of bounded and holomorphic functions on $D$ for which $u_r(W)$ has strong limit as $r\to 1^-$. For $u\in H_W^\infty$ the operator $u(W)$ is defined by
\b\label{n0.a1}
u(W):=\slim_{r\to1^-} u_r(W)\,.
\e

Given a semigroup $\{W_t\}_{t\geq 0}$ of contractions with cogenerator $W$ on a Hilbert space $\H$, one has \cite[Sect.III.8-9]{NAGY-FOIAS}: 
$$
\begin{array}{l}
\ds W_t=e_t(W)\,,\qquad (t\in\R^+)\,,
\\
\ds W=\slim_{t\to0^+} \varphi_t(W_t)\,,
\end{array}
$$
where 
$e_t(\lambda):=\exp\left(t{\lambda+1\over \lambda-1}\right)$ and 
$\varphi_t(\lambda):={\lambda-1+t\over \lambda-1-t}$, for $t\in\R^+$.

\begin{remark}\label{RN2} \rm 
Note that, since $1$ is not an eigenvalue of the cogenerator $W$, every function of $H^\infty$ which is defined and continuous on $\overline{D}\backslash\{1\}$ belongs to the class $H_W^\infty$ \cite[Th.III.2.3]{NAGY-FOIAS}.
This is in particular the case for the functions $e_t$ ($t\geq 0$),
which are holomorphic on the whole complex plane except the point $1$ and satisfy $|e_t(\l)|\leq 1$ on $D$ and $|e_t(\l)|=1$ on $C\backslash\{1\}$.
\end{remark}
 
Intertwining relations can be extended from cogenerators to semigroups and conversely:

\begin{lemma}\label{Np.0}
Let $W$ and $W'$ be contractions on Hilbert spaces $\H$ and $\H'$, respectively, and let  $\LL :\H'\to\H$ be a bounded operator satisfying the intertwining relation
\b\label{n0.a}
W\LL  =\LL  W'.
\e
Then, 
\b\label{n0.b}
u(W)\LL  =\LL  u(W')\,,\quad (u\in H_W^\infty\cap H_{W'}^\infty)\,.
\e
In particular, if $W$ and $W'$ are the cogenerators of the semigroups $\{W_t\}_{t\geq 0}$ and $\{W'_t\}_{t\geq 0}$, respectively, then (\ref{n0.a}) is equivalent to
\b\label{n0.c}
W_t\LL  =\LL  W'_t\,,\quad (t\geq 0)\,.
\e
\end{lemma}

\begin{proof}
From (\ref{n0.a}) it is obvious that 
$W^k\LL  =\LL  W'^k$ for $k=0,1,2,\ldots$
Then, from (\ref{at}),
$$
a(W)\LL  =\LL  a(W')\,,\quad (a\in A)\,,
$$
and (\ref{n0.b}) follows from this and (\ref{n0.a1}).
Now, if $W$ and $W'$ are cogenerators of semigroups of contractions, the functions $e_t$ ($t\geq 0$) belong to $H_W^\infty\cap H_{W'}^\infty$ (Remark \ref{RN2}); therefore, (\ref{n0.c}) follows from (\ref{n0.b}) in this case. Conversely, since $\varphi_t\in H^\infty$ for every $t>0$ and strong limit preserve intertwining relations (involved operators are bounded and there is no problem with domains), (\ref{n0.c}) implies (\ref{n0.a}).
\end{proof}

We are ready to extend Propositions \ref{CN11.1} and \ref{CN11.3} for single operators to groups and semigroups: 

\begin{proof}[of Theorems \ref{TN11.3} and \ref{TN11.1}]
With respect to the minimal dilations of semigroups of contractions and their cogenerators we have the following:
Let $W$ be the cogenerator of a semigroup $\{W_t\}_{t\geq 0}$ of contractions on a Hilbert space $\H$, let $U,\K$ and $U_+,\K_+$ be the minimal unitary and isometric dilations of $W$,  and $R,\RR$ the residual part. Then $U$, and $R$ are the cogenerators of the groups of unitary operators $\{U_t\}_{t\in\R}$ and $\{R_t\}_{t\in\R}$  on $\K$ and $\RR$, respectively, and $U_+$ is the cogenerator of the semigroup $\{U_{+t}\}_{t\geq0}$ of isometries on $\K_+$, where $U_t$, $U_{+t}$ and $R_t$ are the corresponding minimal unitary and isometric dilations and residual part of $W_t$ for each $t\geq0$. See \cite[Sect.III.9]{NAGY-FOIAS} for details.
Theorems \ref{TN11.3} and \ref{TN11.1} are straightforward consequences of these facts, Propositions \ref{CN11.3} and \ref{CN11.1} and Lemma \ref{Np.0}. 
\end{proof}


\section{Functional models}\label{sfm}

Functional models for contractions on separable Hilbert spaces have been given by Sz.-Nagy and Foia\c s \cite{NAGY-FOIAS} on the basis of operator-valued characteristic functions. 
Let $D$ and $C$ denote the open unit disc of the complex plane $\mathbb C$ and its boundary:
$D:=\{\l\in\C:|\l|<1\}$ and $C:=\{\om\in\C:|\om|=1\}$.
In $C$ interpret measurability in the sense of Borel and consider the normalized Lebesgue measure $d\om/(2\pi)$.
Given a separable Hilbert space $\H$, let $L^2(\H)$ denote the set of all measurable functions $v:C\to \G$ such that ${1\over 2\pi}\int_C ||v(\om)||^2_\H\,d\om<\infty$ (modulo sets of measure zero); measurability here can be interpreted either strongly or weakly, which amounts to the same due to the separability of $\H$ \cite{HALMOS61}. The functions in $L^2(\H)$ constitute a Hilbert space with pointwise definition of linear operations and inner product given by 
$(u,v):={1\over 2\pi}\int_C \big(u(\om),v(\om)\big)_\H\,d\om$, ($u,v\in L^2(\H)$).
Let us denote by $H^2(\H)$ the Hardy class of functions
$u(\l)=\sum_{k=0}^\infty \l^k a_k$, ($a_k\in\H$), 
with values in $\H$, holomorphic on $D$, and such that
${1\over 2\pi}\int_C ||u(r\om)||^2_\H\,d\om$, ($0\leq r<1$),
has a bound independent of $r$ or, equivalently, such that $\sum ||a_k||^2_\H<\infty$ \cite[Sect.V.1]{NAGY-FOIAS}.

For a contraction $W$ on $\H$ we can define the {\it defect operators}
$$
D_W:=(I_\H-W^*W)^{1/2}\,,\quad D_{W^*}:=(I_{\H}-WW^*)^{1/2}\,,
$$
which are selfadjoint and bounded by $0$ and $1$, with {\it defect spaces}
$$
\D_W:=\overline{D_W\H}=(\text{Ker }D_W)^\perp\,,\quad
\D_{W^*}:=\overline{D_{W^*}\H}=(\text{Ker }D_{W^*})^\perp\,.
$$
The {\it characteristic function} of $W$,
$$
\Theta_W(\l):=[-W+\l D_{W^*}(I-\l W^*)^{-1} D_W]_{|\D_W}\,,
$$
is defined at least on $D$ where it is a contractive analytic function valued on the set of bounded operators from $\D_W$ into $\D_{W^*}$.
For almost all $\om\in C$ (with respect to the normalized Lebesgue measure)
$\Theta_W(\om):=\slim \Theta_W(\l) $ exists when $\l\in D$ and $\l\to \om$ non-tangentially and
coincides with the previous definition of $\Theta_W(\om)$ when $\om\in A_W$.
In particular we have $\Theta_W(\om)=\slim_{r\to 1^-} \Theta_W(r\om)$ almost everywhere (a.e.) on $C$.
Such function induces a decomposable operator $\Theta_W$ from $L^2(\D_W)$ into $L^2(\D_{W^*})$ defined by
$$
[\Theta_W v](\om):=\Theta_W(\om)v(\om),\quad \text{ for } v\in L^2(\D_W).
$$
The function $\Theta_W$ on $D$ can be recovered from its boundary values on $C$ by means of Cauchy or Poisson integrals (see \cite[Sect.4.7]{RR85} for details). The function $\Theta_W$ is called an {\it outer function} if $\overline{\Theta_W H^2(\D_W)}=H^2(\D_{W^*})$, the adherence taken in $L^2(\D_{W^*})$.
For those $\om\in C$ at which $\Theta_W(\om)$ exists, thus a.e., set
$\Delta_W(\om):=[I-\Theta_W(\om)^*\Theta_W(\om)]^{1/2}$.
$\Delta_W(\om)$ is a selfadjoint operator on $\D_W$ bounded by $0$ and $1$.
As a function of $\om$, $\Delta_W(\om)$ is strongly measurable and generates by
$$
[\Delta_W v](\om):=\Delta_W(\om)v(\om),\quad \text{ for } v\in L^2(\D_W)\,,
$$
a selfadjoint operator $\Delta_W$ on $L^2(\D_W)$ also bounded by $0$ and $1$.

\begin{definition}\rm
A contraction $W$ on $\H$ is called {\it completely non-unitary} ({\it c.n.u.}) if for non-zero reducing subspace $\H_0$ for $W$ is $W_{|\H_0}$ a unitary operator.
\end{definition}

The following canonical decomposition will be useful \cite[Th.I.3.2]{NAGY-FOIAS}:

\begin{lemma}[Sz.-Nagy and Foia\c s]\label{NF.T.I.3.2}
To every contraction $W$ on a Hilbert space $\H$ there corresponds a uniquely determined decomposition of $\H$ into an orthogonal sum of two subspaces reducing $W$, say $\H=\H_0\oplus\H_1$, such that $W^{(0)}=W{|\H_0}$ is unitary and $W^{(1)}=W{|\H_1}$ is c.n.u..
In particular, for an isometry, this canonical decomposition coincides with the Wold decomposition.
\end{lemma}

For a contraction $W$ with decomposition $W=W^{(0)}\oplus W^{(1)}$ as in Lemma \ref{NF.T.I.3.2} one has
$D_W=0\oplus D_{W^{(1)}}$, $D_{W^*}=0\oplus D_{W^{*(1)}}$, $\D_W=\D_{W^{(1)}}$ and $\D_{W^*}=\D_{W^{*(1)}}$.
Hence, $\Theta_W(\l)=\Theta_{W^{(1)}}(\l)$.
Moreover, the canonical decomposition $W=W^{(0)}\oplus W^{(1)}$ for the cogenerator $W$ of a semigroup $\{W_t\}_{t\geq 0}$ induces the same type of decomposition of the semigroup:
$W_t=W_t^{(0)}\oplus W_t^{(1)}$, ($t\geq 0$), 
with $W_t^{(0)}=e_t(W^{(0)})$ unitary and $W_t^{(1)}=e_t(W^{(1)})$ c.n.u.. 
Therefore we pay attention to c.n.u. contractions only.
In order to study the unitary quasi-affine transforms of a contraction $W$ this restriction does not suppose loss of generality because the unitary part of $W$ incorporates itself into any unitary quasi-affine transform of $W$.

A first consequence of our results is a characterization of contractions $W$ with unitary quasi-affine transforms ($W$ c.n.u. or not) in terms of their characteristic function $\Theta_W$:

\begin{proposition}
Let $\H$ be a separable Hilbert space and let $W$ be a contraction on $\H$ such that $\text{Ker}\,W=\{0\}$. The following assertions are equivalent:

i) $W$ has unitary quasi-affine transforms;

ii) the characteristic function $\Theta_W$ of $W$ is outer and 
$$\text{Ker}\,\Theta_W\cap H^2(\D_W)=\{0\}\,.$$  
\end{proposition}

\begin{proof}
Assume without loss of generality that $W$ is a c.n.u. contraction.
It is well known that the following assertions are equivalent: (a) $W$ is of class $C_{\cdot1}$; (b) $\Theta_W$ is an outer function; (c) the operator $\LL_0=P_H|\RR:\RR\to\H$ defined in (\ref{intop}) has dense range. Also the following assertions are equivalent: ($\alpha$) $\LL_0$ is injective; ($\beta$) $\text{Ker}\,\Theta_W\cap H^2(\D_W)=\{0\}$. See \cite[Prop.VI.3.5]{NAGY-FOIAS} and \cite[Prop.2]{KER87} for details. The result follows from these facts and Proposition \ref{CN11.3}.
\end{proof}

\begin{proposition}\label{tfmuqat}
Let $\H$ be a separable Hilbert space and let $W$ be a c.n.u. contraction on $\H$ of class $C_{\cdot1}$ and such that $\text{Ker}\,W=\{0\}$. Every unitary quasi-affine transform of $W$ is unitarily equivalent to the following functional model:
$$
\begin{array}{l}
{\hat \RR}:= \overline{\Delta_W L^2(\D_W)}\,,
\\[1ex]
{\hat R}(v):= \om\, v(\om)\,,\quad (v\in {\hat \RR})\,.
\end{array}
$$
\end{proposition}
$W$ is itself unitarily equivalent to the functional model given by
$$
\begin{array}{l}
{\hat \H}:=[H^2(\D_{W^*})\oplus \overline{\Delta_W L^2(\D_W)}]\ominus \{\Theta_W w\oplus \Delta_W w: w\in H^2(\D_W)\}\,,
\\[1ex]
{\hat W}(u_*\oplus v):=P_{\hat H}(\om u_*(\om)\oplus \om v(\om))\quad (u_*\oplus v\in \hat \H)\,,
\end{array}
$$
where $P_{\hat \H}$ is the orthogonal projection of ${\hat \K}_+:=[H^2(\D_{W^*})\oplus \overline{\Delta_W L^2(\D_W)}]$ onto $\hat \H$.

\begin{proof}
By Proposition \ref{CN11.3} every unitary quasi-affine transform of $W$ is unitarily equivalent to the residual part of its minimal isometric dilation. Then the result follows from the result about functional models for c.n.u. contractions on separable Hilbert spaces, its dilations and residual parts given in \cite[Sect.VI.2 and Th.VI.3.1]{NAGY-FOIAS}.  
\end{proof}

When $W$ is the cogenerator of a semigroup $\{W_t\}_{t\geq0}$, Proposition \ref{tfmuqat} together with Theorem \ref{TN11.1} and Remark \ref{rcd} give a functional model for every unitary quasi-affine transform of $\{W_t\}_{t\geq0}$ and for $\{W_t\}_{t\geq0}$ itself, which are respectively of the form: 
$$
{\hat R}_t(v):= e_t(\om) v(\om)\,,\quad (v\in {\hat \RR},\,t\in\R)\,,
$$
$$
{\hat W}_t(u_*\oplus v):=P_{\hat H}(e_t(\om) u_*(\om)\oplus e_t(\om) v(\om)),\quad (u_*\oplus v\in \hat \H,\,t\geq0)\,.
$$

An alternative form to this model can be given in which the roles of the unit disc $D$ and its boundary $C$ are taken over by the upper half-plane and the real axis \cite{FO64}.
Indeed, for an arbitrary separable Hilbert space $\H$, the spaces $L^2(\H)$ and $H^2(\H)$ are tansformed unitarily --for measures $d\om/(2\pi)$ on $C$ and $dx/\pi$ on $\R$-- onto, respectively, the space $L^2(\R;\H)$ of all (strongly or weakly) Borel measurable functions $f:\R\to \H$ such that ${1\over \pi}\int_\R ||f(x)||^2_\H\,dx<\infty$ (modulo sets of measure zero) and the Hardy class $H^2(\R;\H)$ consisting of the limits on the real axis of the functions $f(z)$ which are analytic on the upper half-plane and for which 
$\sup_{0<y<\infty} \int_\R ||f(x+iy)||_\H^2\,dx<\infty$.
This is carried out by means of the transformation $u\to f$, where
$$
f(x)={1\over x+i}\, u\left({x-i\over x+i}\right).
$$
Then the functional model takes on the following form:
\begin{theorem}\label{tfmuqatg}
Let $\H$ be a separable Hilbert space and let $\{W_t\}_{t\geq 0}$ be a contraction semigroup on $\H$ of class $C_{\cdot,1}$ and with cogenerator $W$  such that $\text{Ker}\,W=\{0\}$.
Then every unitary quasi-affine transform of $\{W_t\}_{t\geq 0}$ is unitarily equivalent to the following functional model:
$$
\begin{array}{l}
{\tilde \RR}:= \overline{\Upsilon_W L^2(\R;\D_W)}\,,
\\[1ex]
{\tilde R}_t(g):= e^{itx} g(x)\,,\quad (g\in {\tilde \RR},\,t\in\R)\,,
\end{array}
$$
where $\ds \Xi_W(z):=\Theta_W\left({z-i\over z+i}\right)$ and 
$\ds \Upsilon_W(x):=[I-\Xi_W(x)^*\Xi_W(x)]^{1/2}$.
\end{theorem}
The semigroup $\{W_t\}_{t\geq 0}$ is itself unitarily equivalent to the functional model
$$
\begin{array}{l}
{\tilde H}:=[H^2(\R;\D_{W^*})\oplus \overline{\Upsilon_W L^2(\R;\D_W)}]\ominus \{\Xi_W w\oplus \Upsilon_W w: w\in H^2(\R;\D_W)\}\,,
\\[1ex]
{\tilde W}_t(f_*\oplus g):=P_{\tilde H}(e^{itx} f_*(x)\oplus e^{itx} g(x)),\quad (f_*\oplus g\in \tilde H,\,t\geq0)\,.
\end{array}
$$


\section{Spectral properties}\label{s5}

The study of the spectral properties of any unitary quasi-affine transform of a contraction $W$ (with $\text{Ker}\,W=\{0\}$), by virtue of Proposition \ref{CN11.1}, reduces to the analysis of the spectrum of the residual part of the minimal dilations of $W$. In addition, the following result is straightforward from Corollary \ref{coro1} together with Theorems 3 and 4 in K\'erchy \cite{KER89}.

\begin{proposition}\label{coro2}
Let $\H$ and $\H'$ be complex Hilbert spaces, $W$ a contraction on $\H$ of class $C_{\cdot1}$ and such that $\text{Ker}\,W=\{0\}$, and $U'$ an arbitrary unitary quasi-affine transform of $W$ acting on $\H'$. Then:

a) if $W$ has an invariant subspace $M$ and the matrix of $W$ with respect to the orthogonal decomposition $\H=M\oplus N$ is 
$$
W=\left(\begin{array}{cc}W_1 &*\\0 & W_2\end{array}\right)\,,
$$
then $U'$ is unitarily equivalent to an orthogonal sum of unitary quasi-affine transforms $U'_1$ and $U'_2$ of $W_1$ and $W_2$, respectively, i.e., the matrix of $U'$ with respect to $\H'=M'\oplus N'$ has the form 
$$
U'=\left(\begin{array}{cc}U'_1 &0\\0 & U'_2\end{array}\right)\,,
$$
where $M'=\bigvee_{n\in\N} U'^{-n}\LL M$.

b) the spectrum $\sigma(W)$ contains $\sigma(U')$ everywhere, that is, $\sigma(W)\supset\sigma(U')$ and for every non-empty-closed-and-open subset $\sigma$ of $\sigma(W)$ one has $\sigma\cap\sigma(U')\neq \emptyset$;

c) the resolvent function of $W$ dominates that of $U'$ in norm:
$$
||(U'-zI)^{-1}||\leq ||(W-zI)^{-1}||\,,\quad (z\in\C\backslash \sigma(W))\,.
$$
\end{proposition}

With respect to the point spectra $\sigma_p(U')$ and $\sigma_p(W)$, i.e. the set of eigen\-va\-lues, from Lemma \ref{N1} and von Neumann mean ergodic theorem \cite{HALMOS} we can deduce the following result in which the intertwining operator $\LL $ is arbitrary and not necessarily a quasi-affinity.

\begin{lemma}\label{N3}
Let $\H$ and $\H'$ be complex Hilbert spaces, let $W$ be a contraction on $\H$ and let $U'$ be a unitary operator on $\H'$.
Assume $\om_0$ to be an eigenvalue of $U'$ and let $u_0$ be a corresponding eigenvector.
Then for every intertwining operator $\LL\in\I(U',W)$
either $\LL  u_0=0$ or $\om_0$ is also an eigenvalue of $W$ and $\LL  u_0$ is a corresponding eigenvector.
\end{lemma}

\begin{proof}
Let $U$ be the minimal unitary dilation of $W$, let $E_U$ and $E_{U'}$ be the spectral measures for $U$ and $U'$, respectively, and let $\LL _+$ the lifting of $\LL $ given in Lemma \ref{N1} (and Remark \ref{RN1}).
According to von Neumann mean ergodic theorem \cite{HALMOS} we get
\b\label{ds1}
\begin{array}{rl}
\LL _+u_0 & \ds =\slim_{n\to\infty} U^n\LL  U'^{-n}u_0
=\slim_{n\to\infty} {1\over N} \sum_{n=0}^{N-1} U^n\LL  U'^{-n}u_0
\\[2ex]
& \ds =\slim_{n\to\infty} {1\over N} \sum_{n=0}^{N-1} \om_0^{-n}U^n\LL  u_0
=E_{U}(\{\om_0\})\LL  u_0,
\end{array}
\e
because $E_{U}(\{\om_0\})$ is just the orthogonal projection over the subspace of $\H$ of vectors invariant for $\om_0^{-1}U$.
Now, from (\ref{n.e}),
\b\label{ds2}
\slim_{n\to\infty} \left[U^{-n}\LL _+-\LL  U'^{-n}\right]=0.
\e
But, since $U$ and $U'$ are unitary, $U\LL _+=\LL _+U'$ is equivalent to $\LL _+U'^{-1}=U^{-1}\LL _+$ and then $\LL _+U'^{-n}=U^{-n}\LL _+$ for $n\in\N$. Thus (\ref{ds2}) coincides with 
\b\label{ds3}
\slim_{n\to\infty} (\LL _+-\LL )U'^{-n}=0.
\e
From (\ref{ds1}) and (\ref{ds3}) we obtain
$$
\slim_{n\to\infty} (\LL _+-\LL )U'^{-n}u_0=\slim_{n\to\infty} \om_0^{-n}(E_U(\{\om_0\})-I_\H)\LL  u_0=0.
$$
Therefore either $\LL  u_0=0$ or $\om_0$ is also an eigenvalue of $U$ and $\LL  u_0$ is a co\-rres\-pon\-ding eigenvector.
From this we obtain the result since the eigenvalues of modulus $1$ and its corresponding eigenvalues coincide for $W$ and $U$ \cite[Prop.II.6.1]{NAGY-FOIAS}.
\end{proof}

An immediate consequence of Lemma \ref{N3} is the following:

\begin{proposition}\label{ps}
Let $\H$ be a complex Hilbert spaces and let $W$ be a contraction on $\H$ of class $C_{\cdot1}$ and such that $\text{Ker}\,W=\{0\}$. Then for every unitary quasi-affine transform $U'$ of $W$ one has
$$
\sigma_p(U')=\sigma_p(W)\cap C\,.
$$
\end{proposition}

\begin{proof}
Lemma \ref{N3} implies $\sigma_p(U')\subseteq\sigma_p(W)\cap C$, since now $\LL$ is a quasi-affinity and $\LL u_0\neq0$ for $u_0\neq 0$. On the other hand, the eigenspace corresponding to each $\l\in \sigma_p(W)\cap C$ is in the unitary part of $W$ (Lemma \ref{NF.T.I.3.2}), which takes part of $U'$, so that $\sigma_p(U')\supseteq\sigma_p(W)\cap C$.
\end{proof}

The analysis of the spectrum of the residual part can be carried out through the functional model of $W$ (Section \ref{sfm}) when we restrict attention to c.n.u. contractions on complex separable Hilbert spaces.

\begin{definition}\rm
For a c.n.u. contraction $W$, let $\varepsilon(W)$ denote the set of points $\om\in C$ at which $\Theta_W(\om)$ exists and is not isometric. For any subset $\alpha$ of $C$, the {\it essential support}, denoted by ``ess supp $\alpha$", is defined as the complement with respect to $C$ of the maximal open subset of $C$ whose intersection with $\alpha$ is of zero Lebesgue measure.
\end{definition}

\begin{proposition}\label{SNS}
Let $\H$ be a complex separable Hilbert spaces and let $W$ be a contraction on $\H$ of class $C_{\cdot1}$ and such that $\text{Ker}\,W=\{0\}$. Then the spectrum $\sigma(U')$ of every unitary quasi-affine transform $U'$ of $W$ is the perfect set (closed set without isolated points) of $C$
$$
\sigma(U')=\text{ess supp }\varepsilon(W).
$$
Moreover, the spectrum of $U'$ is absolutely continuous, $\sigma(U')=\sigma_{ac}(U')$. Then, $\sigma_p(W)\cap C=\sigma_p(U')=\emptyset$.
\end{proposition}

\begin{proof}
The result follows from Proposition \ref{CN11.1} and \cite[Prop.VII.5.1]{NAGY-FOIAS}.
That $\sigma_p(W)=\sigma_p(U')=\emptyset$ follows from Proposition \ref{ps}.
\end{proof}

In the light of Propositions \ref{ps} and \ref{SNS}, the point spectra $\sigma_p(U')$ and $\sigma_p(W)$, and the continuous singular spectrum $\sigma_{sc}(U')$, are associated to the unitary part of $W$ only.
More interesting results about the spectrum of the residual part and its multiplicity have been obtained by Petrov \cite{PE93} and Exner-Jung \cite{EJ98}.

Results in this Section extend without difficulty to semigroups of contractions and their unitary quasi-affine transforms. Moreover, {\it Foia\c s-Mlak spectral mapping theorem} \cite{FM66} states that for a c.n.u. contraction $W$ the spectral mapping theorem holds in the usual sense, i.e. 
$$
\sigma[\mu(W)]=\mu[\sigma(W)]\,,
$$ 
if the set of points of $C$ to which the function $\mu\in H^\infty$ can be continuously extendable include all $\om\in\sigma(W)\cap C$.
Thus, under the conditions of Theorem \ref{SNS}, if $W$ and $U'$ are cogenerators of 
a semigroup $\{W_t\}_{t\geq 0}$ of c.n.u. contractions on $\H$ and a  group $\{U'_t\}_{t\in\R}$ of unitary operators on $\H'$, respectively, and 
$1\notin\sigma(W)$, then 
$$
\begin{array}{ll}
\sigma[W_t]=e_t[\sigma(W)]\,, & (t\geq 0)\,,
\\[2ex]
\sigma[U'_t]=e_t[\sigma(U')]\,, &(t\in\R)\,.
\end{array}
$$

\section{Intrinsic randomness in statistical mechanics}\label{sirsm}

By way of conclusion let us apply the above results to MPC theory in the context of statistical mechanics. For it, consider an abstract dynamical system $(\Omega,{\mathcal A},\mu,\{S_t\})$, where $\Omega$ is the phase space of the system equipped with the 
$\sigma$-algebra ${\mathcal A}$ and $\{S_t\}$ is a group of measurable point transformations on $\Omega$ preserving the probability measure $\mu$ (automorphic case). As it has been commented in the Introduction, the evolution of density functions (states) $\rho$ in $L^2=L^2(\Omega,{\mathcal A},\mu)$  under the given deterministic dynamics is described by the Frobenius-Perron unitary group $\{U_t\}$ induced by $\{S_t\}$:
$$
U_t\rho(\om):= \rho(S_{-t}\om)\,,\qquad  \om\in\Omega\,.
$$

On the other hand, every Markov process on $\Omega$ with stationary distribution $\mu$ is associated with a continuous semigroup of contractions $\{W_t\}$ on $L^2$ preserving positivity (i.e. $\rho\geq0$ implies $W_t\rho\geq0$ for $t\geq0$) and satisfying $W_t1=1$, where the constant density $1$ is the equilibrium state. From the point of view of the second law of thermodynamics, we are interested only in irreversible Markov processes which correspond to monotonic Markov semigroups, i.e.
\b\label{mdm}
||W_t\rho-1|| \text{ decreases monotonically to } 0 \text{ as } t\to\infty\,,
\e
for all states $\rho\neq 1$. 

\begin{definition}\rm
The deterministic dynamics with induced unitary group $\{U_t\}$ on $L^2$ is said to be {\it intrinsically random} \cite{M78} if there exists a quasi-affinity $\Lambda\in\I(U_t,W_t)$ for $t\geq0$ (see Definition \ref{dqss}) for a monotonic Markov semigroup $\{W_t\}$. In such case, $\{W_t\}$ is called a {\it random image} of $\{U_t\}$.
This implies that $\Lambda$ preserves positivity, $\Lambda 1=1$, and $\Lambda$ preserves normalization (i.e. $\int\rho\,d\mu=\int \Lambda\rho\,d\mu$, for $\rho\geq0$).  
\end{definition}

In what follows we focus attention on flows (continuous time parameter $t$) and assume $\text{Ker}\,W=\{0\}$ for the cogenerator $W$ of $\{{W_t}\}$ (equivalently, $-1\notin \sigma_p(A)$ for the generator $A$ of $\{{W_t}\}$).
The superfluous one-dimensional subspace of $L^2$ spanned by the constant functions shall be denoted by $\C$.

Among other things, we have proved the following:

\begin{corollary}
Let $\{U_t\}_{t\in\R}$ be an intrinsically random unitary dynamics with random  image $\{W_t\}_{t\geq0}$ on $L^2$. Then:
\begin{itemize}
\item[a)]
$\{{W_t}_{|(L^2\ominus\C)}\}_{t\geq0}$ is a c.n.u. contraction semigroup of class $C_{01}$, i.e., 
$$
\begin{array}{l}
\slim_{t\to\infty} {W_t}_{|(L^2\ominus\C)}=0 \text { and }
\\
\slim {W_t}_{|(L^2\ominus\C)}^*\rho\neq 0 \text{ for every nonzero }\rho\in L^2\ominus\C\,;
\end{array}
$$ 
\item[b)]
$\{U_t\}_{t\in\R}$ is unitarily equivalent to the residual group of the minimal dilation of $\{W_t\}_{t\geq0}$;
\item[c)]
$\{U_t\}_{t\in\R}$ is unitarily equivalent to the group of unitary $*$-asymptotes of $\{W_t\}_{t\geq0}$, in particular, those of Remark \ref{ruaa};
\item[d)]
$\{{U_t}_{|(L^2\ominus\C)}\}_{t\in\R}$ is unitarily equivalent to the functional model given in Theorem \ref{tfmuqatg}, with $W$ the cogenerator of $\{{W_t}_{|(L^2\ominus\C)}\}_{t\geq0}$;
\item[e)]
$\sigma_p(U_t)=\sigma_p(W_t)\cap C=\{1\}$ and the eigenspace is $\C$, for every $t\geq0$;
\item[f)]
$\sigma({U_t}_{|(L^2\ominus\C)})=\sigma_{ac}({U_t}_{|(L^2\ominus\C)})= \text{ess supp }\varepsilon(W_t)$, for every $t\geq 0$.
\end{itemize}
\end{corollary}

\begin{proof}
(a) follows from Theorem \ref{TN11.3} and condition (\ref{mdm}).
(b) is just Theorem \ref{TN11.1}. (c) is a consequence of Corollary \ref{coro1}. (d) follows form Theorem \ref{tfmuqatg}. (e) and (f) follow from Proposition \ref{SNS}.  
\end{proof}

\section*{Acknowledgements}
The author wishes to thank Professor Z. Suchanecki for useful discussions and the staff of Facult\'e des Sciences, de la Technologie et de la Communication of Universit\'e du Luxembourg for kind hospitality. 
This work was supported by JCyL-project VA013C05 (Castilla y Le\'on) and MEC-project FIS2005-03989 (Spain).


\appendix

\section{Similarity}\label{as1}

The relation of {\it similarity} between operators has been defined in Definition \ref{dqss}.
For contraction operators, conditions for similarity to unitary operators have been given by Sz.-Nagy-Foia\c s \cite[Sect.IX.1]{NAGY-FOIAS} and Gokhberg-Kre\u\i n \cite{GK67}.
A study of similarity to unitary operators for more general classes of operators can be found in 
Sakhnovi\v c \cite{Sakh68}, van Casteren \cite{vC83} and Naboko \cite{Nabo84}. 

For the sake of completeness, the criteria obtained by Sz.-Nagy-Foia\c s \and Gokhberg-Kre\u\i n  are collected in the following:  

\begin{proposition}\label{NF.C.IX.1.4}
Let $W$ be a contraction operator on a Hilbert space $\H$. The following statements are equivalent:
\begin{itemize}
\item[a)]
$W$ is similar to a unitary operator.
\item[b)]
$W$ is invertible and $\lim_{n\to\infty}||W^{-n}||<\infty$.
\item[c)]
The strong limit $V:=\slim_{n\to\infty}{W^*}^{n}W^{n}$ (which exists and is a non-negative operator) is uniformly positive, that is $V\geq \delta I$ for some positive number $\delta$.
\item[d)]
$W$ is invertible and $\slim_{n\to\infty}{W^*}^{-n}W^{-n}$ exists and belongs to $\L(\H)$.
\item[e)]
The open unit disc $D$ is contained in the resolvent set of $W$ and there exists a constant $a$ such that
$$
||(\l I-W)^{-1}||\leq {a\over 1-|\l|}\,,\quad (\l\in D)\,.
$$
Actually, it suffices to assume that at least one point of $D$ belongs to the resolvent set of $W$ and that
$$
|(\l I-W)h||\geq {1-|\l|\over a}||h||\,,\quad (\l\in D,\, h\in\H)\,.
$$
\item[f)]
The characteristic function $\Theta_W$ of $W$ satisfies the conditions
$$
||\Theta_W(\l)g||\geq c||g||\,,\quad (\l\in D,\, g\in\D_W)\,,
$$
and
$$
\Theta_W(\l)\D_W=\D_{W^*}\,,\quad (\l\in D)\,,
$$
the second one at least at one (and then at every) point of $D$.
\item[g)]
$\Theta_W(\l)$ is boundedly invertible at every $\l\in D$ and $||\Theta_W(\l)^{-1}||$ have a bound independent of $\l$ on $D$.
\item[h)]
There exists a left continuous decomposition of the identity $P(t)$ $(0\leq t\leq 2\pi)$ separating
the spectrum $\sigma(W)\subseteq C$ of $W$ (i.e. (i) $P(t)WP(t)=WP(t) (0\leq t\leq 2\pi)$, (ii) $\sigma(W|P(\tau)\H)\subset\{e^{it}\colon 0\leq t\leq\tau\}$ and (iii) $\sigma((I-P(\tau))W|(I-P(\tau))\H)\subset\{e^{it}\colon\tau\leq t\leq 2\pi\}$)
and such that the operator function $D_W^{1/2}P(t)D_W^{1/2}$ satisfies a Lipschitz condition, that is
$$
||D_W^{1/2}(P(t_2)-P(t_1))D_W^{1/2}||\leq C\,|t_2-t_1|\,,\quad (0\leq t_1<t_2\leq2\pi)\,.
$$
\end{itemize}
\end{proposition}

Under the conditions of Proposition \ref{NF.C.IX.1.4}, $W$ is similar in particular to the residual part of its minimal unitary dilation. Moreover, the least upper bound of $||\Theta_W(\l)^{-1}||$ on $D$ is equal to the minimum of $||\LL||$ and $||\LL^{-1}||$ for the intertwining operators $\LL$ such that $\LL W\LL^{-1}$ is unitary.


\end{document}